\documentstyle{article}
\include{graphicx}

\include{natbib}
\include{amsmath,amsfonts,amssymb}
\newcommand{\be}{\begin{equation}}
\newcommand{\ee}{\end{equation}}
\newcommand{\bea}{\begin{eqnarray}}
\newcommand{\eea}{\end{eqnarray}}

\begin{document}

\title{Peaks in the CMBR power spectrum. II. Physical interpretation for any cosmological scenario}

\author{Mart\'\i n L\'opez-Corredoira\\
Instituto de Astrof\'\i sica de Canarias\\
C/.V\'\i a L\'actea, s/n\\
E-38200 La Laguna, Tenerife, Spain,\\
AND\\
Departamento de Astrof\'\i sica,\\
Universidad de La Laguna,\\
E-38206 La Laguna, Tenerife, Spain\\
E-mail: martinlc@iac.es}

\maketitle

{\bf \large ABSTRACT}

In a previous paper (part I), the mathematical properties of the cosmic microwave 
background radiation power spectrum which presents oscillations were discussed. Here, 
we discuss the physical interpretation: a power spectrum with 
oscillations is a rather normal characteristic expected from any fluid with 
clouds of overdensities that emits/absorb radiation or interact gravitationally 
with the photons, and with a finite range of sizes and distances for those clouds. 
The standard cosmological interpretation of ``acoustic'' peaks
is just a particular case; peaks in the power spectrum might be generated in 
scenarios within some alternative cosmological model that have nothing to do 
with oscillations due to gravitational compression in a fluid.

We also calculate the angular correlation function of the anisotropies from the 
WMAP-7yr and ACT data, in an attempt to derive the minimum number of parameters a 
polynomial function should have to fit it: a set of polynomial functions with a total
of $\approx$6 free parameters, apart from the amplitude, is enough to reproduce 
the first two peaks. However, the standard model with six tunable free parameters 
also reproduces higher order peaks, giving  the standard model a higher confidence. 
At present, while no simple function with six free parameters is found to give 
a fit as good as the one given by the standard cosmological model, we may
consider the predictive power of the standard model beyond an instrumentalist 
approach (such as the Ptolemaic astronomy model of the orbits of the planets).



\section{Introduction}

Analysis of the Cosmic Microwave Background Radiation (CMBR) is one of the most important pieces of observational information supporting the standard model of cosmology, together with the abundance of light elements, the distribution of galaxies and their redshift--distance calibration, and other data. Moreover, in the last two decades, the CMBR has also become a source of accurate values for the parameters of this standard model of cosmology, which is usually called ``precision Cosmology''.\cite{Pri05} The foundations of cosmology are thought to be definitively established, and now a quantitative science is pursued for which the fitting of the power spectrum of the distribution of CMBR anisotropies and few other data gives us the numerical values of the parameters in the equations governing the entire past, present and future of our Universe.

The history of CMBR analysis is one of both successes and failures in the confirmation of  predictions (that is, theoretical calculations carried out {\it before} the observations are available) and plenty of {\it ad hoc} fittings and corrections of the standard model. For instance, the fact that the CMBR presents an isotropic black body shape was a successful prediction. However, the first predictions of the (antenna) temperature of the CMBR were wrong: none of the predictions of the background temperature, which ranged between 5 K and 50 K,\cite{Van93} matched observations, the worst being  the value of 50 K given by Gamow in 1961; in 1965, the year of the discovery of the correct temperature, a value of 30 K was calculated by Dicke et al.\cite{Dic65} from the amount of helium production observed. The first predictions of CMBR anisotropies were also wrong, The first  values predicted being one part in hundred or thousand;\cite{Sac67} however, this value could not fit the observations, which gave values hundreds of times smaller, so non-baryonic dark matter was introduced {\it ad hoc} to solve the question.

CMBR analyses in the last decade have concentrated on anisotropies on small angular scales, smaller than the angular resolution of several degrees achieved by the  COBE-DMR in the '90s. The power spectrum of the anisotropies of BOOMERANG and MAXIMA-1 showed a first peak at the Legendre multipole 
$\ell \approx 200$,\cite{deB00,Han00} 
corresponding to angular scales of $\approx 1.2^\circ $, which was interpreted as a discovery of the previously predicted acoustic peaks in a flat Universe with 
$\Omega =\Omega _m+\Omega _\Lambda =1$, with the greatest contribution from a dark energy component of $\Omega _\Lambda =0.7$ (see review in Ref. \cite{Hu02} and references therein for further comments). Indeed, there are two parts in the comparison 
%of observations 
of theoretical predictions and observations: a successful one, and a half-successful/half-failure one. The totally successful prediction was the qualitative shape of the power spectrum containing several peaks: For instance, Peebles \& Yu\cite{Pee70} had predicted a CMBR power spectrum with fluctuations, the acoustic peaks, from the hypothesis of primeval adiabatic perturbations in an expanding universe. However, the position of the first peak was not at the expected position when first measured: already in the mid-90s the position of the first peak was determined to be $\ell \approx 200$  with other experiments to measure small angular resolution anisotropies previous to BOOMERANG or MAXIMA-1 but with smaller sky coverage (Ref. \cite{Whi96} and references therein). De Bernardis et al.\cite{deB00} and Hanany et al.\cite{Han00} just added a refinement in the measurement of the position of the first peak but they were not its discoverers. White et al.\cite{Whi96} realized that the preferred standard model of the time (an open Universe with 
$\Omega =\Omega _m\approx 0.2$ and without dark energy) did not fit the observations, so that they needed a larger $\Omega $. This was one of the elements, together with SNIa observations, which would encourage cosmologists to include the new {\it ad hoc} element:  dark energy. Between 1997 and 2000 this change of mentality in standard cosmology was produced, and then, in 2000, with the results of the new BOOMERANG and MAXIMA-1 experiments, cosmologists were proud to announce that new observations were giving exactly the results they expected. In any case, just paying attention to the first analyzes of the first peak in the mid-90s, we attribute a half-success to the prediction because, even though they failed to fit the observations with the preferred standard model of a curved open Universe at that time, the idea of a flat Universe was also presented as a possibility 
in the '90s, and indeed a possibility preferred by inflationary paradigms.
The position of the other peaks would also serve to constrain the cosmological models. The acoustic peaks on angular scales of 1 deg and 0.3 deg were predicted with the second peak nearly as high as the first one in $\ell (\ell +1)C_\ell/(2\pi )$.\cite{Bon87,Jor95} Later, as data became available, its amplitude was reduced and the positions of the other peaks and their relative heights were also constrained from the model with a higher agreement with the data (e.g., Ref. \cite{Hu01}).

More recently, the analysis of CMBR polarization anisotropies (e.g., Ref. \cite{Lar11}) has also provided strong support for the standard cosmology. This indicates the way in which the CMBR is polarized due to the scattering of free electrons. Photon diffusion into regions of different temperatures are possible only when the plasma becomes sufficiently optically thin; these diffused photons could then scatter only while there are still free electrons left. Since photons could not diffuse too far, polarization cannot vary much over very large angular scales.

Given this history of CMBR analyses, it is difficult to accept that 
our observations reproduce ``predictions''. Nonetheless, whichever comes first,  theory or  observation, the fact remains that we have now have a cosmological model which is able to fit  the CMBR power spectrum $C_\ell$ quite accurately. It is usually thought that this may not happen by chance, given the apparent complexity of the power spectrum shape, whereas the six-parameter cosmological model is relatively simple.\cite{Lar11,Pei05} In order to fit the temperature--polarization cross power spectrum and the polarization--polarization power spectrum,\cite{Lar11} one would need an extra parameter (optical depth), so a total of at least seven free parameters are necessary.
In this paper, we will not talk about the polarization anisotropies and their fits with a seven-parameter model. Roughly speaking, the relationship between temperature-temperature and the polarization-temperature, or polarization-polarization power spectra is expected since they are different ways of seeing the same light with different filters; a full discussion of the generation of polarization is not the subject of this article.

Some critics of the standard model\cite{Dis07,Lop09} claim that there is little value in the fact that the standard cosmological model might fit some observational cosmological data if the number of free parameters in the model is comparable to the number of independent parameters  characterizing the observations. For instance, Ptolemaic geocentric astronomy may fit the observations of the orbits of the planets but with too high a number of free parameters in the theory. 
The principle of Occam's Razor tells us that a theory is better when the number of free parameters is low. Occam's Razor can indeed be understood with a Bayesian analysis, in which it is tested how probable  a model  is with a given number of free parameters with respect to other
models with a higher number of free parameters.\cite{Jef92,Ber92} Otherwise, if a theory has a high number of free parameters, it loses credibility because it is always possible to create a ``false'' model to fit some data when the number of free parameters is comparable to the number of degrees of freedom in the data.
Philosophers of science, when talking about cosmology,  associate this approach with ``instrumentalism'' (e.g., Ref. \cite{Sol08}).
And saying that the power spectrum contains hundreds of independent parameters for a given resolution is not correct, because the different values of $C_\ell $ for each $\ell$ are not independent in the same sense that hundreds of observations of the position and velocity of a planet
do not indicate hundreds of independent parameters,  the information on the orbit of planet being instead reduced to only six Keplerian parameters. 

For the reasons just given, understanding how much information is in the power spectrum is important for key questions in the discussion of the fundamentals of cosmology, and this is the matter we shall discuss in this paper. In spite of the huge number of papers dedicated to analyzing how to fit the power spectrum with  standard cosmological models, there is a dearth of literature concerning the properties of this power spectrum independently of the standard physical model. Indeed, there are here two main points which should be clarified:
\begin{enumerate}
\item Are the oscillations something atypical in a power spectrum? That is, should we consider the fact the power spectrum contains oscillations a successful prediction of the standard cosmological model which cannot be produced by any other means? 
\item How much information is contained in the power spectrum? That is, how many free parameters in a function are necessary to fit the CMBR power spectrum? Should we consider the fitting of the power spectrum with a model of six free parameters as a validation of the standard cosmological model? 
\end{enumerate}

Our paper is structured as follows. In \S \ref{.oscill}, we discuss
the physical origin of the oscillating peaks in the power
spectrum, and of the corresponding scales in the angular
correlation function, which corresponds to the physical interpretation
of the mathematical toy models developed by L\'opez-Corredoira \& Gabrielli\cite{Lop13}(hereafter Paper I). 
In \S \ref{.corr}, we analyze the angular correlation function
of CMBR anisotropies on small scales, and we find
that a set of polynomial functions with a total of $\approx$6 free
parameters may  roughly fit the angular correlation function. Finally,
in \S \ref{.discuss} we will discuss these results and summarize them
in \S \ref{.concl}.

\section{The physical origin of oscillations in the power spectrum}
\label{.oscill}

The fluctuation field  on the
sky, $\frac{\delta T}{T}(\theta ,\phi )$ can be decomposed into
spherical harmonics in the sphere:
\begin{equation}
\frac{\delta T}{T}(\theta ,\phi )=\sum _{\ell=0}^\infty \sum
_{m=-\ell}^{+\ell}a_{\ell m}Y_{\ell m}(\theta ,\phi ) \;.
\end{equation}
Assuming statistical isotropy, the  
angular power spectrum $C_\ell$ is 
defined as
\begin{equation}
C_\ell \equiv \langle |a_{\ell m}|^2\rangle=\frac{1}{2\ell +1}
\sum _{m=-\ell }^{+\ell }|a_{\ell m}|^2 \;, 
\end{equation}
whereas the angular two-point correlation function
is defined as  (Paper I):
\be
C(\theta )\equiv \left \langle \frac{\delta T}{T}(\theta
_1,\phi _1)\frac{\delta T}{T}(\theta _2,\phi _2)\right \rangle 
,\ee
where $\theta =\theta _2-\theta _1$.  
The two-point correlation function can be expanded in 
terms of Legendre polynomials 
\be
C(\theta ) = \frac{1}{4\pi }\sum _{\ell =0}^\infty (2\ell+1)C_\ell
P_\ell [\cos (\theta )] \;,
\label{ctheta}
\ee
where $P_\ell $ are the Legendre polynomials. 
By inverting Eq. (\ref{ctheta}) we find\cite{Cop10}
\begin{equation}
C_\ell =2\pi \int _0^\pi C(\theta )P_\ell [\cos (\theta )]\sin (\theta )d\theta .
\label{cl}
\end{equation}

A remarkable advantage of the use of self-correlation rather than the
power spectrum is that it gives more direct information on the
physical space (Paper I), so we shall use it here.

\subsection{CMBR anisotropies in terms of the Toy Models of Paper I}

In Paper I, it was shown that oscillations observed in $C_\ell $
could be generated by a self-correlation $C(\theta )$ with an abrupt transition
for some angle. That is, for an angular correlation function such that
\begin{equation}
\label{abrupt}
C(\theta )=\left \{ 
\begin{array}{ll}
g_1(\theta ),& \mbox{$\theta \le \theta_0$}
 \\
g_2(\theta ),& \mbox{$\theta > \theta_0$}
\end{array}
\right \} \;, 
\end{equation}
with $g_1^{(n)}(\theta_0^-)\ne g_2^{(n)}(\theta_0^+)$ and $n$ some integer number greater than
or equal to one; that is, a function with some derivative which is not continuous at 
$\theta _0$. Also in Paper I, we saw that a toy model in which 
we distribute disks of fixed or variable radii with some distribution of 
temperature $T$, but with the value of the field outside
the disks being $T=0$, we get angular correlation of the type of Eq. (\ref{abrupt}) and
consequently we get oscillations in the power spectrum.
We wonder now whether these toy models of Paper I 
could produce a CMBR distribution which mimicked the observed power spectrum/angular correlation, and what their physical interpretation would be.

We could interpret the disks as projections of spherical clouds in 3D space that emit/absorb radiation or interact gravitationally with it.
Toy models 1 of Paper I restrict the angular radius of the disks/clouds to be a constant.
The case of toy model 1b, with the restriction that disks do not intersect in their projection, 
produces a dip for $2R<\theta _0<2R+\delta $ for some small $\delta $. 
Other examples of matter distribution over spherical bubbles 
that do not produce oscillations are also possible (e.g., Ref. \cite{Col02}, \S 13.4). 

\begin{figure}
\vspace{1cm}
{\par\centering \resizebox*{5.cm}{5.cm}{\includegraphics{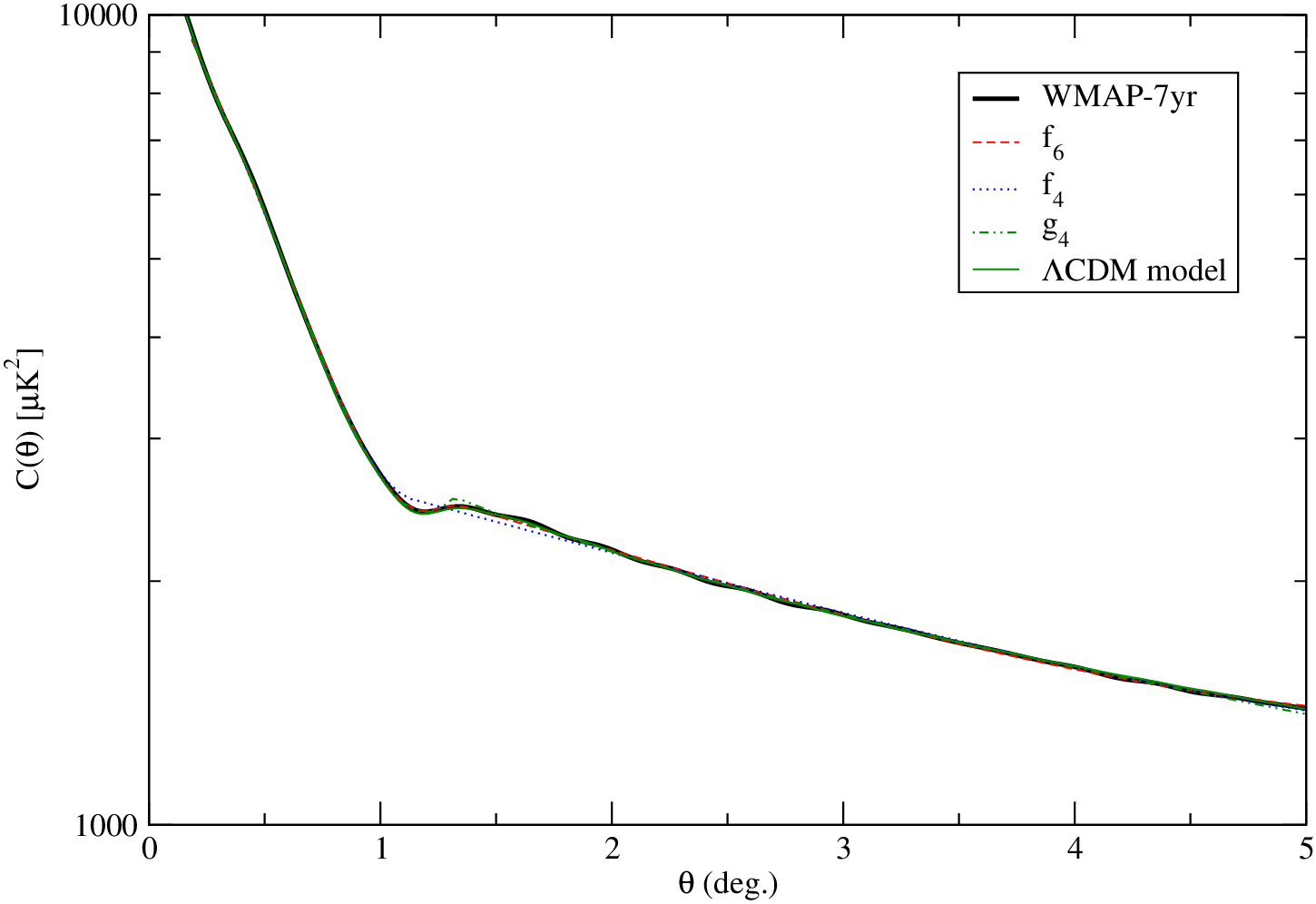}}\par}
\vspace{.6cm}
{\par\centering \resizebox*{5.cm}{5.cm}{\includegraphics{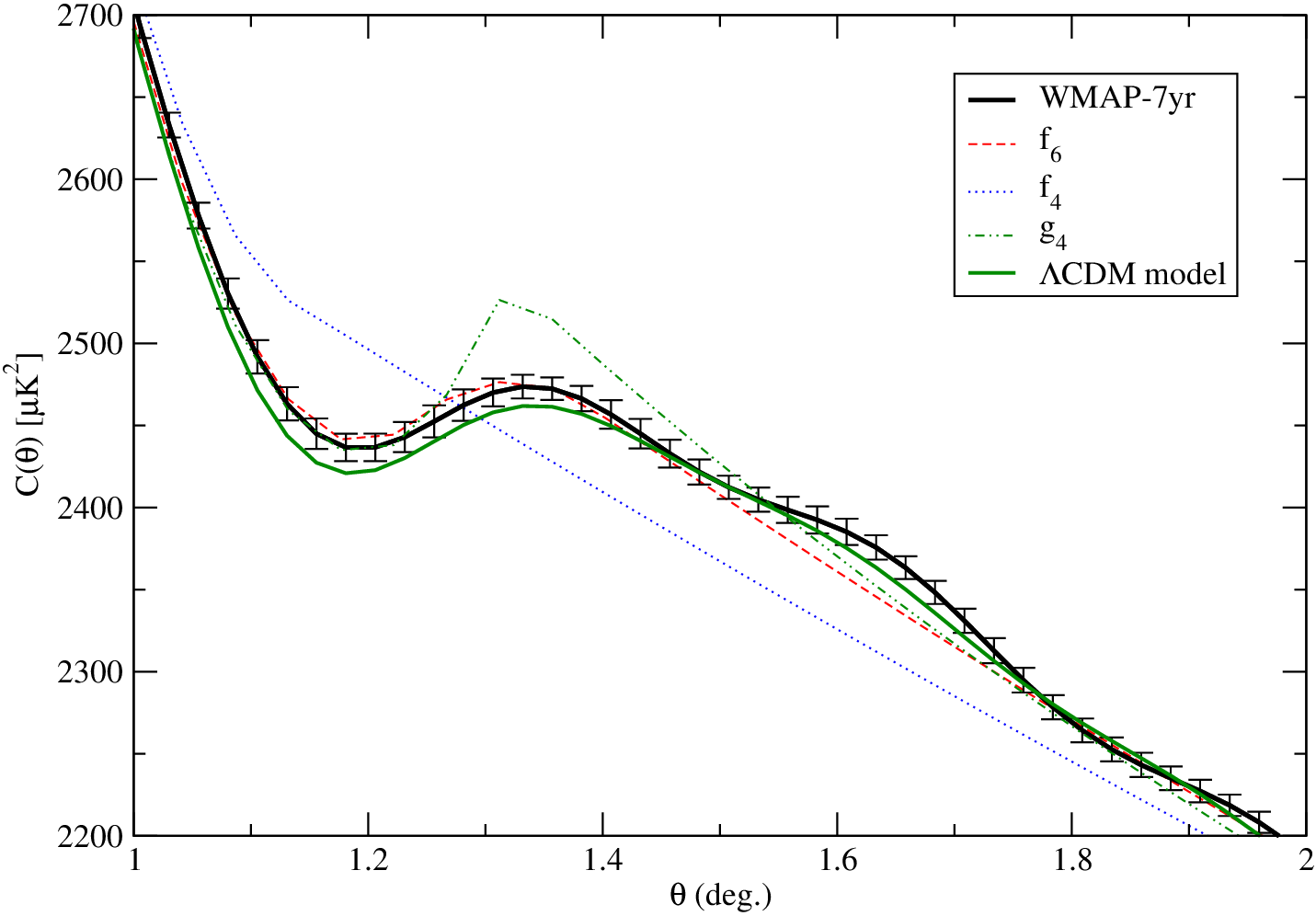}}\par}
\vspace{.6cm}
{\par\centering \resizebox*{5.cm}{5.cm}{\includegraphics{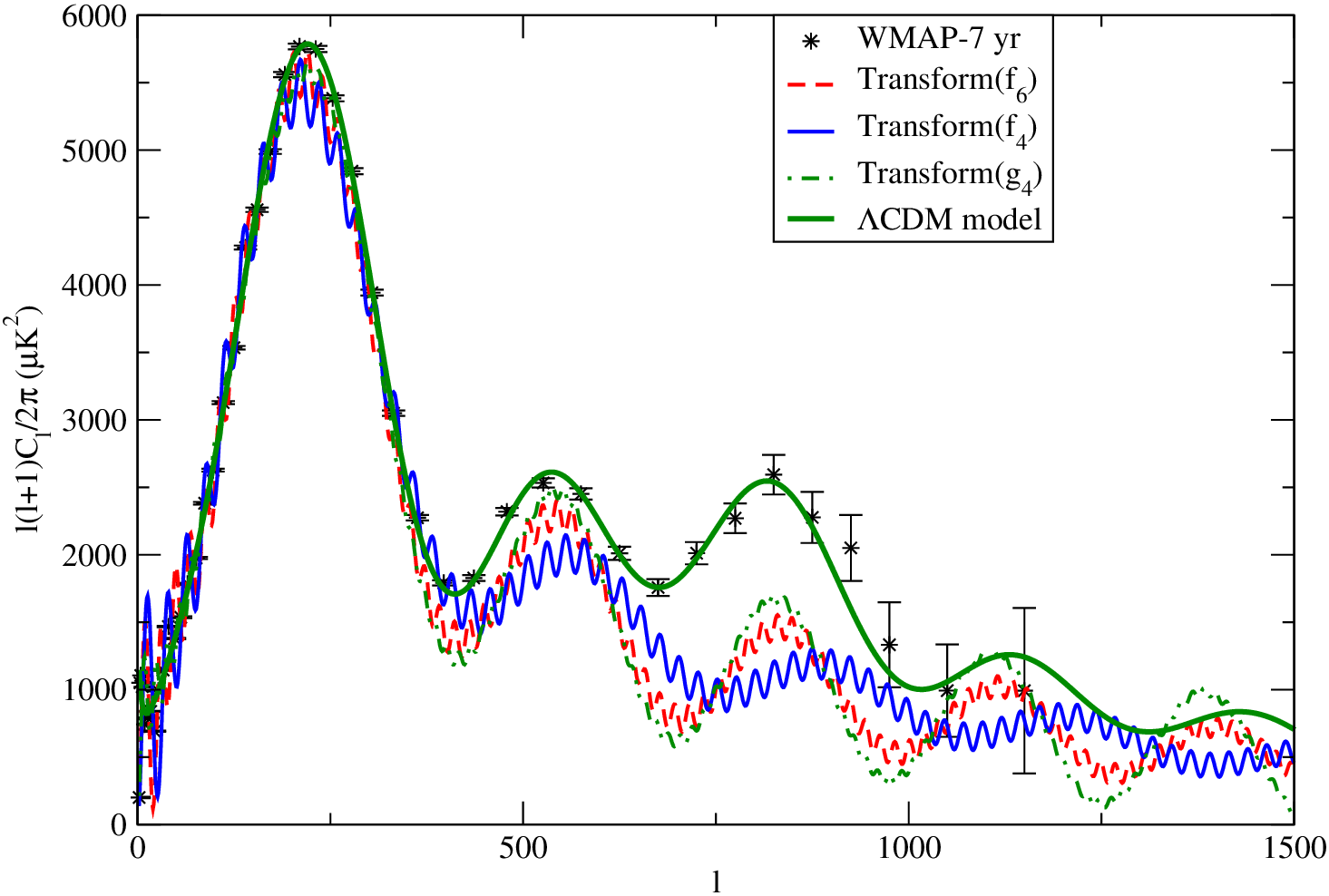}}\par}
\caption{
Top/Middle: Self-correlation function of WMAP-7yr CMBR
  angular anisotropies and fits to it. Errors of WMAP correlation only
  include instrumental noise. Bottom: Power spectrum of these fits,
  together with the WMAP-7yr data. Also included is the best fit
  with the $\Lambda $CDM mode and its
  angular correlation (plus a constant term $-401\mu $K$^2$ in order to
  compensate for the effects of  cosmic variance at low $\ell$ in the
  data).}
\label{Fig:corrWMAP}
\end{figure}

Just looking at Fig. \ref{Fig:corrWMAP}/top, we can see there are two clear
regimes for $C(\theta )$ in the real data with a transition at $\theta \approx
1.2^\circ $, which is directly comparable with the toy models of Paper I. 
A value of $R\approx 0.6^\circ $ would be
necessary to assimilate it to the results of the toy model 1
(Fig. 3 of Paper I).  The case of toy model 1b, with the
restriction that disks do not intersect in their projection, produces
a dip for $2R<\theta <2R+\delta $ for some small $\delta $, and this
is indeed observed in the real CMBR self-correlation too (see Fig. \ref{Fig:corrWMAP}/Middle). From this
analysis, we get the idea that with a simple model of a distribution
of clouds of constant physical radius at a fixed distance, with some
anticorrelation among them, which emit/absorb radiation or interact
gravitationally with the radiation, we can get the CMBR
self-correlation, its corresponding power spectrum with the
oscillations, or even with a distribution of clouds of variable radii
and distances (toy model 2).

The standard model obeys the scenario of the toy model 1 of Paper I: The redshift associated with the  matter--radiation decoupling gives us a fixed distance and the constant physical size is given by the size of the acoustic horizon. The origin of oscillations in the power spectrum is well understood in terms
of a photon--baryon fluid compressed by gravitational attraction produced by local density fluctuations giving rise
 to acoustic oscillations.\cite{Pee70},\cite{Gab05}(\S 6.6.2) The angular size $\theta =1.2^\circ $ corresponds to the acoustic horizon size at recombination.\cite{Bas02,Her04}
In any case, apart from the standard model, any physical scenario resembling the conditions of
the toy models 1 may produce a power spectrum with the observed features. 

If we were to think about non-standard cosmologies, considering a distribution of disks with a contant radius at constant distance would be very restrictive. But toy models 2 of Paper I teach us that a constant radius and distance are indeed unnecessary for those clouds to generate an abrupt 
transition. The case of toy model 2a shows us that a distribution of variable sizes of the clouds still produces abrupt transitions
at the minimum ($2R_{min}$) and maximum ($2R_{max}$) angular sizes of the cloud. The case of toy model 2b shows us that,
for a fixed linear size ($L$), and a distribution of clouds in 3D space in a range of distances between $r_{min}$ and
$r_{max}$, we get abrupt transitions in the correlation function at $2L/r_{max}$ and $2L/r_{min}$. A combination of variable
sizes and variable distance is also possible with similar conclusions. A distribution of clouds over an infinite distance, apart from the problem of Olber's paradox, would produce a value of $C(\theta )$ that is continuous in all of its derivatives and that would not give rise to any oscillation.

The conclusion to take from the analysis of these toy models, therefore, is
that we could in principle generate a distribution of CMBR anisotropies with oscillations in the
power spectrum by setting a fluid with some clouds of overdensities
(with respect to the average density), which emit/absorb radiation or interact gravitationally with the radiation over a finite range of sizes and distances.

The question which now arises is how much fine tuning we need to fit a model following
the mathematical recipes of the toy models in Paper I onto the observed CMBR anisotropy
distribution. Indeed, the toy models of Paper I have some tunable parameters and
functions $f(\theta )$, $\omega (\theta )$ or $A(R)$, whose variation can generate the different $C(\theta )$ and $C_\ell $ with oscillations:
\begin{equation}
C(\theta )=\int _{R_{\rm min}}^{R_{\rm max}}A(R)h\left(\theta \frac{R_0}{R}\right)dR
\label{toy2}
,\end{equation}\[
h(\theta )=h[f(\theta ),\omega (\theta)]
,\]
where $h(\theta )$ is the correlation function derived from  toy model 1 for disks of angular radius $R_0$, 
and its integration for different angular radii with amplitudes $A(R)$ gives rise to  toy model 2. The 
function $f(\theta )$ gives the isotropic distribution of antenna temperature within each disk. The function
$\omega (\theta )$ gives the two-point correlation function in the distribution of the centers of the disks.

One could produce a distribution
of disks as in toy models 1 and 2 to produce such fits.
Theoretically, a fit of $C(\theta )$ through the toy-model representation would get the best functions $f(\theta )$, $\omega (\theta )$ or $A(R)$, or the different families of functions if there is degeneracy.
From Eq. (\ref{toy2}), it looks generally possible to recover  functions $A$, $f$, $\omega $, which can produce an observed $C(\theta )$. Moreover, if we fix a toy model 1 $h(\theta )$, then, in principle, we 
could find a function $A(R)$ to fit the data by inverting Eq. (\ref{toy2}). However, in such a case this function $A(R)$ might
be very complex and difficult to parameterize with few parameters, so a free combination of the
three functions should be allowed to get the fit with a minimum number of parameters.

Nonetheless, in practice this fit cannot be carried out directly. Testing all the different combinations of functions would be a very time-consuming computer calculation: for each set of $f$, $\omega $, the calculation
of $h$ takes $\sim$10 minutes for $\theta <5^\circ $ in normal computers with the resolution given in Paper I,
and if we go to much higher $\theta $ (required for $\frac{R_0}{R}\ll 1$; the computing time for a given $\theta $ is larger for larger $\theta $) the time is increased to hours for each
$h$, so exploring the multiple possible combinations of $f$ and $\omega $ shape functions would require an excessive time, even with a super-computer. Instead, for the question of the number of necessary parameters, we will analyze in the following section the fit of $C(\theta )$ by a generic polynomial function. 

\section{Parameterization of CMBR angular correlation function}
\label{.corr}

Our problem is to determine the amount of independent information contained in
$C(\theta )$, or its Fourier transform $P(k)$ or $C_l$. The functional dependence is
simpler for the real-space, $C(\theta )$, than in  frequency space (paper I), so
we will concentrate on the analysis of the former.

In signal processing of multidimensional signals, for example in computer vision, the
intrinsic dimension of the signal describes how many variables are needed to represent the signal. For a signal of $N$ variables, its intrinsic dimension is an integer number $M\le N$.
The intrinsic topological dimensionality of  data refers to the minimum number of
free parameters needed to generate a data set.\cite{Jai88}
This concept of ``intrinsic dimension'' can also be applied to our problem: knowing
the function with a minimum number of free parameters which fits our $C(\theta )$.
We want to know the intrinsic dimension of $C(\theta )$. 
It contains hundreds of values for different $\theta $, for the 
given resolution, but they are not independent.
However, although the problem
is defined in a precise way, the ways of finding the solution are not. There
is some literature on algorithms to get the intrinsic dimension on data structures
in high-dimensional space (e.g., Refs. \cite{Keg02,Ban08}), but these are not applicable
to our specific problem.

In this paper, we use fits to simple polynomial functions as a way to estimate
a number of free parameters, but the real intrinsic dimension of $C(\theta )$ might
be lower with other more complex functions.
Certainly, there may exist another
non-trivial function shape which produces better fits.
But we prefer not to explore all the possible known functional shapes 
because this would reduce the probability to being a chance fit. That is, the larger
the number of test functions that we use, the higher the probability of obtaining a better fit, but this probability should be divided by the number of total test functions used in the trials.

\subsection{WMAP angular correlation function}

The direct valculation of the angular correlation function from  CMBR data is subject to
several difficulties.\cite{Her04} Rather, we will carry out the calculation of
the self-correlation of WMAP (Wilkinson Microwave Anisotropy Probe)-7yr on small scales corresponding to the transform [Eq. (\ref{ctheta})] of the published power spectrum.\cite{Lar11}
The result is shown in Fig. \ref{Fig:corrWMAP},
which illustrates the combined power spectrum using all the available frequencies.
Indeed, when we talk about data from observations, $C_\ell ^{obs.}=W_\ell C_\ell ^{real}$, with $W_\ell$ as a window function, usually with
a shape $W_\ell =exp[-\ell (\ell+1)\sigma ^2]$, with $\sigma $ related to the size of the beam. Throughout this paper we shall concentrate on
the ``real'' power spectrum, once corrected for the window function, so from now on $C_\ell \equiv C_\ell^{real}$, and the correlation
function $C(\theta )$ will also refer to the real correlation function rather than that directly obtained from observations. We also assume that foreground contamination has been correctly subtracted, regardless of doubts that may be entertained concerning this subtraction.\cite{Lop07}
The error bars stem from the error in the power spectrum due to the measurement (error derived from diagonal terms of Fisher matrix; the multipole moments are slightly coupled, so a correct treatment of errors would require use of the entire Fisher matrix; small, correlated contribution due to beam and point source subtraction uncertainty are not included; cosmic variance error is not included), calculated through Monte Carlo simulations. The total errors are expected to be slightly larger but not much, as we will see below.

We would like to determine 
how many free parameters are necessary to describe
$C(\theta )$ in Fig. \ref{Fig:corrWMAP}. 
In Fig. \ref{Fig:corrWMAP}, there are best weighted fits of this correlation
function within $0.2^\circ <\theta <5^\circ $ with two functions 
 of class $C^1$ (the function and its first derivative are
continuous) respectively with 4 and 6 free parameters (apart from the
amplitude): $f_4$ and $f_6$, which are respectively consecutive
polynomials of second degree for two or three different angular ranges
satisfying the continuity of $f_i$ and $f_i'$.
\begin{equation}
\label{f6}
f_6[\theta (deg.)](\mu K^2)=
\end{equation}\[
\left \{ 
\begin{array}{ll}
12067-16118\theta +6746\theta ^2,& \mbox{$\theta <1.239$}
 \\
2452+593.4\theta _1-3542\theta _1^2  ,& \mbox{$1.239\le \theta \le 1.391$}
 \\
2460-485.1\theta _2+53.24\theta _2^2,& \mbox{$\theta > 1.391$}
\end{array}
\right \}
,\]\[
\theta _1\equiv \theta-1.239;\ \theta _2\equiv \theta-1.391
\]
[Fit with $N=192$ points giving $\chi ^2=584$].

Note that all of the numbers given in  Eq. (\ref{f6}) are not independent. In Eq. (\ref{f6}), three numbers for each of the three second-degree polynomials are given, and two angles where the transitions are produced, so there are eleven numbers. But imposing continuity on the function and its first derivative, reduces  the number of independent parameters to four (two for each transition point). The amplitude was not included among the free parameters, so in total we have six free parameters in $f_6$.
\begin{equation}
f_4[\theta (deg.)](\mu K^2)=
\end{equation}\[
\left \{ 
\begin{array}{ll}
12144-16434\theta +7012\theta ^2,& \mbox{$\theta <1.140$}
 \\
2523-446.7\theta _3+39.29\theta _3^2,& \mbox{$\theta > 1.140$}
\end{array}
\right \}
,\]\[
\theta _3\equiv \theta-1.140
\]
[Fit with $N=192$ points giving $\chi ^2=1844$].

Another function of class $C^0$ (continuity in the function but not in
its derivative) with four free parameters which  roughly fits $C(\theta )$
is:
\begin{equation}
g_4[\theta (deg.)](\mu K^2)=
\end{equation}\[
\left \{ 
\begin{array}{ll}
12073-16144\theta +6759\theta ^2,& \mbox{$\theta <1.320$}
 \\
2539-877.4[ln(\theta )-ln(1.320)],& \mbox{$\theta > 1.320$}
\end{array}
\right \}
\]
[Fit with $N=192$ points giving $\chi ^2=768$].

We plot these fits in Fig. \ref{Fig:corrWMAP}/top-middle.
We also plot the fit given by the standard $\Lambda$-CDM model,
plus a constant term $C_K$ in order to compensate for the effects of  cosmic variance at low $\ell$ in the data. It is this term that minimizes 
the chi-square in the fit: $C_K=-401\ \mu $K$^2$, $\chi ^2=330$ in the same
$N=192$ points. The fact that we obtain $\chi ^2/N\approx 1.7$ instead of one with the standard model  indicates that the error bars may be underestimated 
by $\approx$25\%.

The fits with $f_6$ and $g_4$ are better than that with $f_4$; they
fit  the small dip in the transition at $\theta \approx
1.2^\circ $ better; in any case, $f_4$ also gives a fairly acceptable fit.
Bashinsky \& Bertshinger\cite{Bas02} showed that, if this dip is cut out, the
oscillations in the power spectrum after the first peak would be
removed; but, as we have shown, this depends on how the dip is
cut out.  Other minor oscillations of the correlation function are
most probably fluctuations due to noise.

The power spectrum associated with the fitted correlations $f_6$, $f_4$
and $g_4$ (for $f_4$ and $f_6$, over 5 degrees, we have extrapolated
linearly till the value of the function reaches null value, and zero
beyond), derived through Eq. (\ref{cl}) applied to $C=f_6$, $C=f_4$,
$C=g_4$ respectively, is plotted in Fig. \ref{Fig:corrWMAP}/bottom. The 
agreement with the observed power spectrum in the
first two peaks is acceptable. All of them reproduce the first peak, and the second
peak is more or less well fitted for $f_6$ or $g_4$. The amplitude of the third peak, however, is not well
fitted, except for its center and width.

\begin{figure}
\vspace{1cm}
{\par\centering \resizebox*{5.5cm}{5.5cm}{\includegraphics{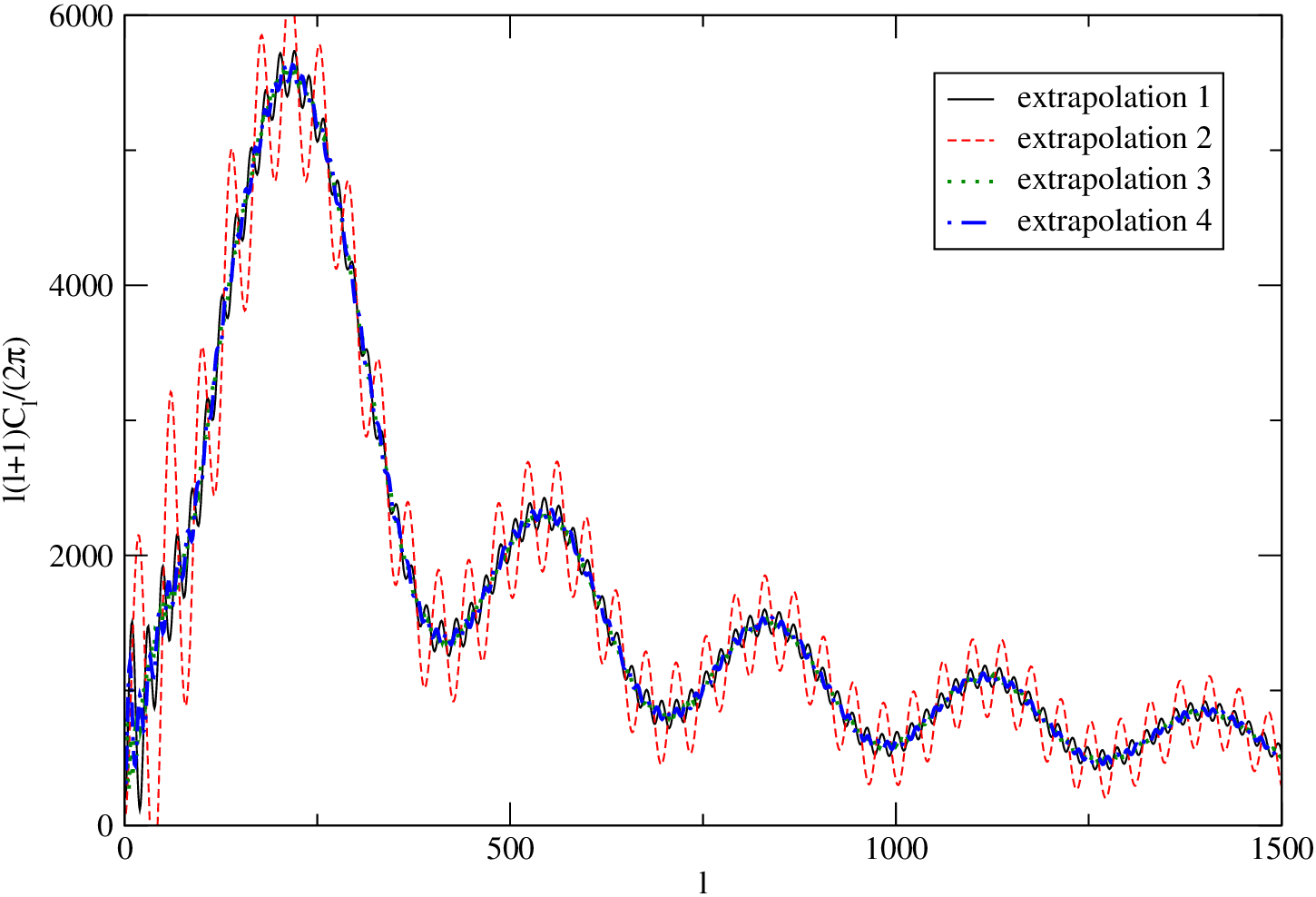}}\par}
\caption{Power spectrum (Fourier transform) of $f_6$ with different extrapolations 
for $5^\circ <\theta <\theta _*$ such that the extrapolation 
reaches $f_6(\theta _*)=0$ and we set $f_6(\theta >\theta _*)=0$:
1) Extrapolation 1 (the linear one used for  
Fig. \ref{Fig:corrWMAP}/down): $f_6 [\theta (^\circ)>5]=f_6(5)+f_6'(5)\times 
(\theta -5)$; 2) Extrapolation 2: $f_6[\theta (^\circ)>5]=f_6(5)+\frac{1}{2}f_6'(5)\times 
[(\theta -4)^2-1]$; 3) Extrapolation 3: $f_6[\theta (^\circ)>5]=f_6(5)+2f_6'(5)\times 
[\sqrt{\theta -4}-1]$; 4) Extrapolation 4: $f_6[\theta (^\circ)>5]=f_6(5)+2000\times \left[\exp \left[\frac{1}{2000}f_6'(5)\times (\theta-5)\right]-1\right]$.}
\label{Fig:extrapolations}
\end{figure}

The minor oscillations in the power spectra of Fig. \ref{Fig:corrWMAP}/bottom 
are produced by the method of extrapolation. We can perform different extrapolations
of the same $f_6$ for $5^\circ <\theta <\theta _*$, all of them maintaining a continuous function
and derivative in $\theta =5^\circ $, such that the extrapolation 
reaches $f_6(\theta _*)=0$ and we set $f_6(\theta >\theta _*)=0$:
1) Extrapolation 1 (the linear one used for  
Fig. \ref{Fig:corrWMAP}/down): $f_6 [\theta (^\circ)>5]=f_6(5)+f_6'(5)\times 
(\theta -5)$; 2) Extrapolation 2: $f_6[\theta (^\circ)>5]=f_6(5)+\frac{1}{2}f_6'(5)\times 
[(\theta -4)^2-1]$; 3) Extrapolation 3: $f_6[\theta (^\circ)>5]=f_6(5)+2f_6'(5)\times 
[\sqrt{\theta -4}-1]$; 4) Extrapolation 4: $f_6[\theta (^\circ)>5]=f_6(5)+2000\times \left[\exp \left[\frac{1}{2000}f_6'(5)\times (\theta-5)\right]-1\right]$. 
Fig. \ref{Fig:extrapolations} shows 
these four cases. As  can be seen, the result of the inversion is the same, and 
the change of extrapolation only affects to the oscillations of high frequency.

Although the power spectrum seems to contain a lot of information with
all their oscillations, indeed all the information is concentrated in
the abrupt transition of the self-correlation function at scales of
1.1-1.4 degrees. In our examples, the second derivatives of $f_6$ are
not continuous at $\theta=1.239^\circ $ and $\theta=1.391^\circ $, the
second derivative of $f_4$ is not continuous at $\theta=1.140^\circ $,
and the first derivative of $g_4$ is not continuous at $\theta
=1.320^\circ $. A simple function with six free parameters is able to
reproduce more or less the correlation. Even with four free parameters, a
rough fit is obtained, especially for $g_4$. 
The best fit with the
standard cosmological model\cite{Lar11} (see
Fig. \ref{Fig:corrWMAP}/down) is indubitably better. The last WMAP
data of the 7-yr release have precisely allowed a better determination
of this third peak,\cite{Lar11} being almost with the same
amplitude as the second peak, and this gives an important advantage
to the standard cosmology with respect to a generic polynomial fit,
which tends to reduce the power of the peaks in their consecutive
sequence. Hu et al.\cite{Hu01} also showed with BOOMERANG and MAXIMA-1 data that
the height of the second and third peaks are very similar. 
The best value of $\chi ^2/N$ for our fits was 3.0 
for $f_6$, which indicates that the standard model ($\chi ^2/N=1.7$) produces
a better fit.

The fact that we have found a polynomial fit such as $f_6$ with six parameters giving
an approximate fit does not mean that this is the best  possible fit with generic
functions with $\le 6$ free parameters. The intrinsic dimension, might be 
6 or even lower and get a much better fit with other kinds of functions.

\subsection{WMAP+ACT angular correlation function}

\begin{figure}
\vspace{1cm}
{\par\centering \resizebox*{5.5cm}{5.5cm}{\includegraphics{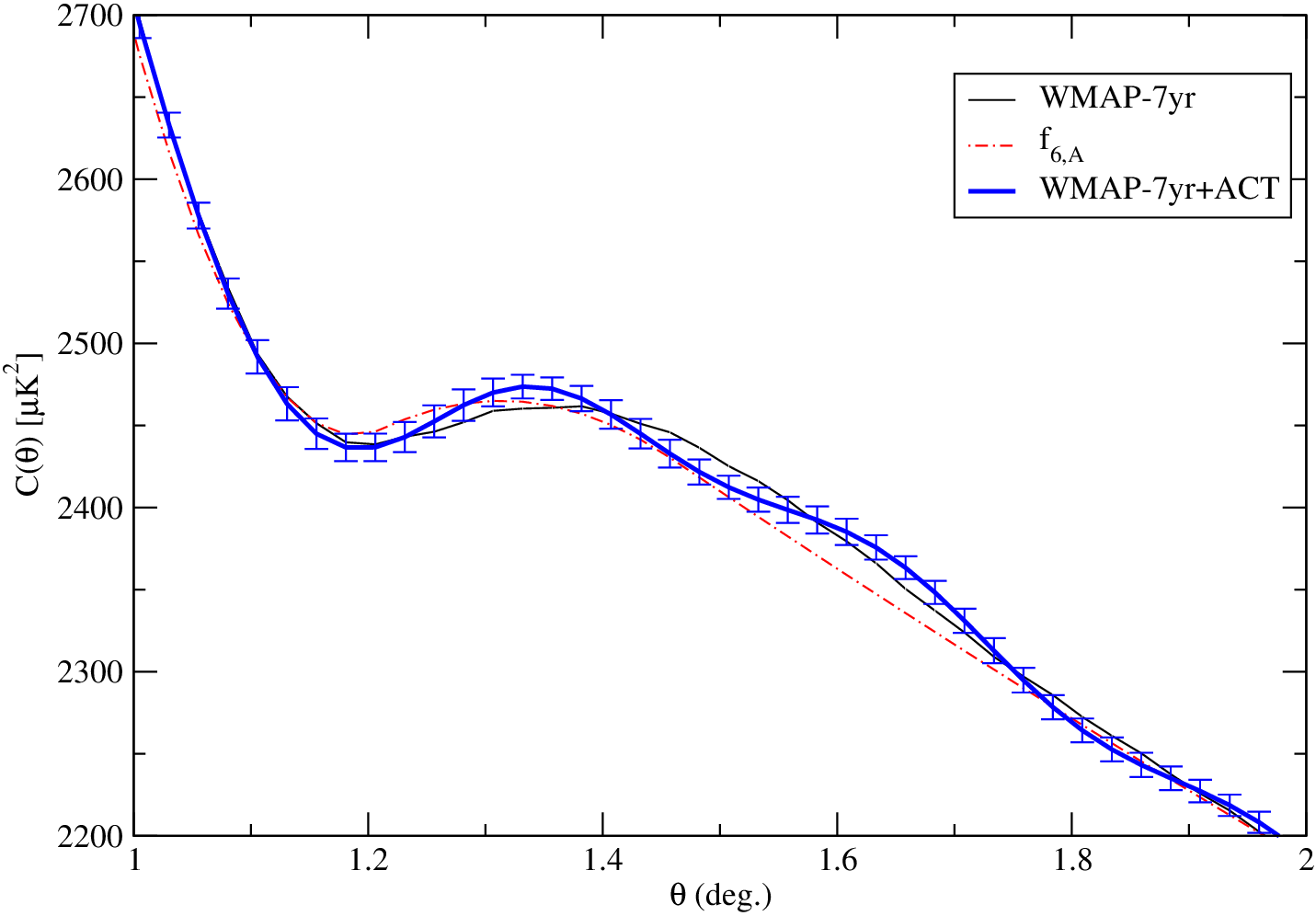}}\par}
\vspace{1cm}
{\par\centering \resizebox*{5.5cm}{5.5cm}{\includegraphics{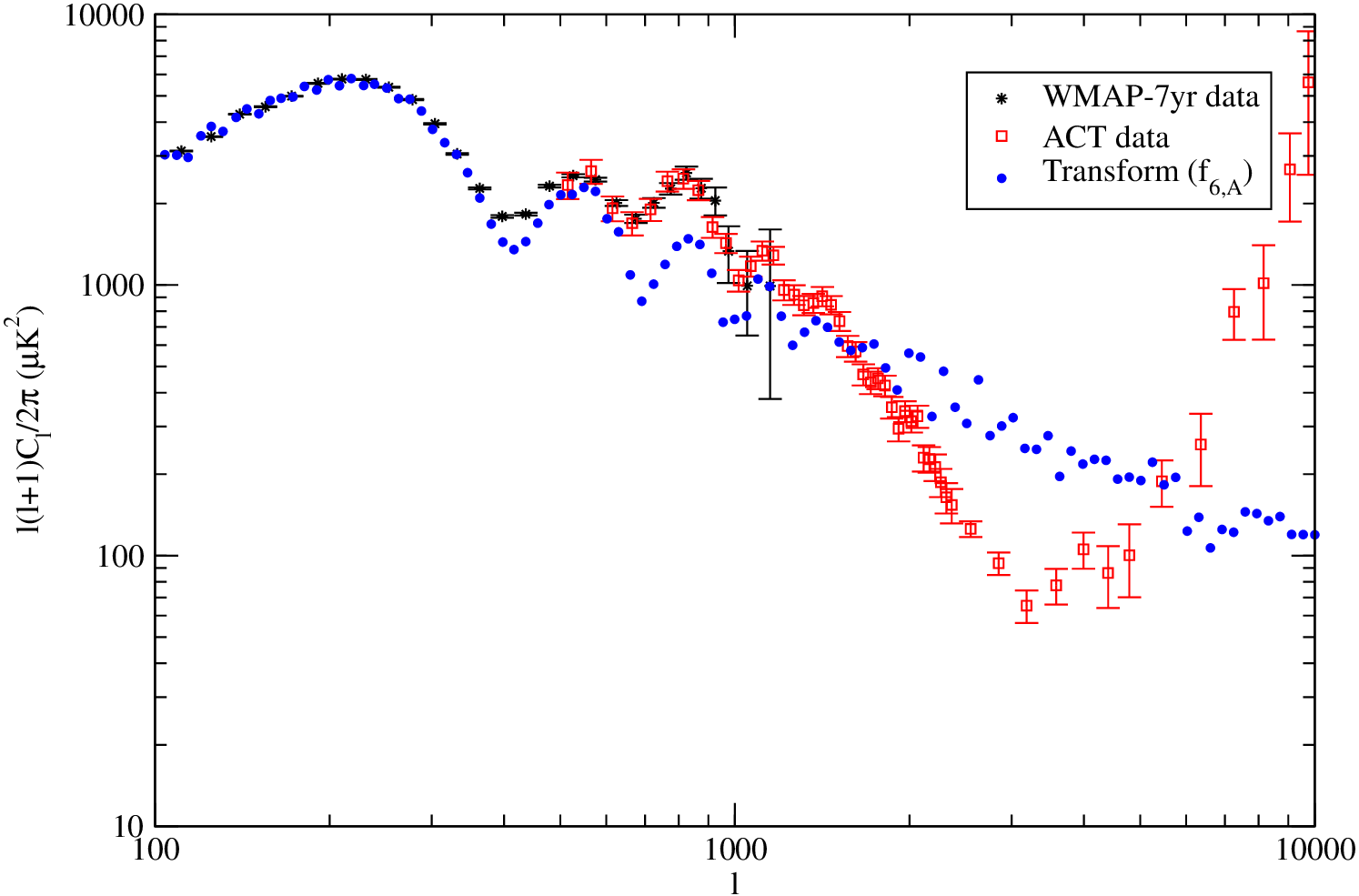}}\par}
\caption{Top: Self-correlation function of WMAP-7yr+ACT CMBR
  angular anisotropies and fits. Bottom: Power spectrum of this fit,
  together with the WMAP-7yr and ACT data.}
\label{Fig:corrACT}
\end{figure}

Higher-order peaks were obtained in other experiments, for instance in the  ACT (Atacama Cosmology Telescope\cite{Das11}) project, and these data might be used for an analysis like the one carried out in the previous subsection. Here, I will carry out the analysis with the 148 GHz data once corrected for the window function, that is: $C_\ell ^{real}=\frac{C_\ell ^{obs.}}{W_\ell }$
for $500<\ell <9750$ (using the available points at http://lambda.gsfc.nasa.gov/product/act/act\_spectra\_get\-.cfm and with linear interpolation for the rest); for $\ell \le 500$ we will use the WMAP data given in the previous subsection. Figure \ref{Fig:corrACT} plots the self-correlation of this WMAP+ACT power spectrum at small scales, corresponding to its transform given by Eq. (\ref{ctheta}). We also use Monte Carlo simulations to estimate the error bars from 
the errors given by the analytical expression from Ref. \cite{Das11}. The results are very similar to the WMAP self-correlation.

Again, a best-weighted fit of this correlation
function is carried out, within $0.2^\circ <\theta <5^\circ $ with a function 
of class $C^1$ and six free parameters, which is a consecutive
polynomial of second degree for three different angular ranges
satisfying the continuity of $f_i$ and $f_i'$:
\begin{equation}
f_{6,A}[\theta (deg.)](\mu K^2)=
\end{equation}\[
\left \{ 
\begin{array}{ll}
12159-16349\theta +6878\theta ^2,& \mbox{$\theta <1.213$}
 \\
2448+335.6\theta _1-1672\theta _1^2  ,& \mbox{$1.213\le \theta \le 1.457$}
 \\
2430-481.9\theta _2+54.12\theta _2^2,& \mbox{$\theta > 1.457$}
\end{array}
\right \}
,\]\[
\theta _1\equiv \theta-1.213;\ \theta _2\equiv \theta-1.457
\]
[Fit with $N=192$ points giving $\chi ^2=260$].

This fit is much better than the equivalent one with WMAP data only, with $\chi ^2/N=1.35$.
Nonetheless, a new transform of this correlation function to obtain the power spectrum (see Fig. 
\ref{Fig:corrACT}/below) has the same problem as with the WMAP data only. The third  and
subsequent peaks are more or less fitted in position but not in amplitude. Again, the standard model
gives a much better fit.\cite{Das11} Note, however, the ACT works at 148 and 218 GHz, from which a fitting model needs to include lensed CMB, Sunyaev-Zel'dovich, and foreground contributions with a total of ten or more free parameters. The intergalactic medium might become dominant on 
$\ell >\sim 2000$.\cite{Atr06}

\section{Discussion}
\label{.discuss}

Apart from the standard model, any physical scenario resembling the conditions of
the toy models 1 or 2 of Paper I, 
with some sets of parameters and functions $f(\theta )$, $\omega (\theta )$ or $A(R)$ which fit the correlation of Fig. \ref{Fig:corrWMAP}, 
may produce a power spectrum with the observed features. Nevertheless, we
must bear in mind that any possible alternative scenario should also explain the almost perfect black body shape of that radiation, and the value of around 3 K in its antenna 
temperature, for which there are exotic ideas in the literature, but which are
not free from caveats and inconsistencies. 
There are, for instance,  proposals  in terms of a local origin of CMBR and its thermalization of stellar radiation by the intergalactic medium in the form of thermalization produced by an aether,\cite{Clu80} carbono needles (``whiskers''\cite{Hoy93,Hoy94}), cosmic meteoroids (``cosmoids''\cite{Sob01}), radiation of electrons orbiting protons in the atoms [against quantum mechanics],\cite{Shp02}
or electrons or ions upon the action of magnetic fields.\cite{Ler88,Ler95,Bry04,Kri09,Cra11}
The origin must be very local in order to preserve the black-body shape, or to be non-redshifted, which seems problematic.
Some proposals have even placed the origin of the CMBR within our own Galaxy, in the local bubble within 100 pc,\cite{Nar10} but it
is puzzling to understand by this means the Sunyaev-Zel'dovich imprint of clusters of galaxies in the CMBR \cite{Agh11} or any possible correlation of the CMBR with  
galaxies.\footnote{This has indeed not been found yet, since the measurements of the cross-correlation of WMAP with galaxy surveys are not significant; see Ref. \cite{Lop10} and references therein.}

Another relevant feature of CMBR anisotropies is the gaussianity of its fluctuations.
However, this should not be a major problem for alternative cosmologies since many
different origins of the fluctuations are expected to be Gaussian, although not all of them. Theories involving inflation generally predict a pattern of Gaussian noise, whereas theories based on symmetry breaking and the generation of defects have more distinctive signatures.\cite{Cou94} Moreover, it is not totally clear yet whether the CMBR fluctuations are exactly Gaussian: A non-Gaussian distribution has been
claimed by several authors (e.g., Ref. \cite{Lop07}, \S 2.3 and references therein; 
Refs. \cite{Rae07,Ber07,McE08,Ros09}); however, other analyses\cite{Rub06,McE06} claim that only a few regions have such non-Gaussian anisotropies due to contamination,  e.g. the Corona Borealis supercluster region, and most of the regions in the sky are Gaussian. It has also been argued\cite{Liu05,Toj06} that the non-gaussianity is associated with cold spots of unsubtracted foregrounds; even the lowest spherical harmonic modes, which should be the cleanest, in the map are significantly contaminated with foreground radiation.\cite{Chi07}

As said in \S \ref{.corr},  
a simple function with six free parameters is able more or less to reproduce  the first two peaks of
the power spectrum. Even with four free-parameters, $g_4$, a reasonable fit is obtained. Hence, one would expect that any theory of the type of toy models 1 or 2 with this number of free parameters could represent the data.
The standard cosmological model fits the power spectrum with six free 
parameters.\footnote{Plus many other parameters which introduce second-order changes. And, even so, there is a degeneracy in the solutions with different values of $H_0$ and $\Omega _\Lambda $: WMAP data and the large scale structure of galaxies could be reproduced without explicitly requesting the existence of dark energy,\cite{Bla03} i.e. with $\Lambda =0$. This degeneracy is broken by adding cosmological information from other sources, for instance, from SNIa data.}\cite{Pei05,Lar11} This is astonishing but it would be much more surprising if it could fit it without free parameters or with only one or two free parameters fitted to the power spectrum. Certainly, the same parameters which are used to fit the CMBR power spectrum are able to
fit other cosmological data, but more than six parameters are necessary, as well as additional
information such as initial conditions, conditions of stellar formation, galaxy formation, how dark matter distributes in galaxies, etc. A global analysis of the cosmological models would require the examination of all of the available independent sources of cosmological
data (nucleosynthesis, supernovae, gravitational lensing, etc.), and to check whether they are comparable to the number of free parameters in the model. We do not carry out this analysis in the present paper. At present, we  refer only to the CMBR anisotropies.

Observe that we are not claiming here that $f_6$ or any best fit could substitute the
standard theory in order to explain the observed $C_\ell $ or $C(\theta )$.
The function $f_6$ is just a simple fit which does not represent any physical model.
The toy models of Paper I may represent a physical model, but not the polynomial functions of \S \ref{.corr}.
A simple fitting recipe cannot substitute a physical theory even if it
requires a smaller number of parameters to reproduce
experimental/observational data.  A physical theory can be substituted
only by another theory which is able to explain all the phenomena
taken into account by the first one and requires fewer subtleties and
{\it ad hoc} assumptions. This is nothing else that the application of the principle of Occam's razor,
 but this is valid for homogeneous things, i.e. two theories
(or physical explanations) and two fitting procedures (for instance
the comparison between a fit with a sum of exponentials or a sum
of power laws) and not a theory and a fitting procedure that have
different intrinsic value. 
Nonetheless, our aim here was just to show the capacity that a wrong
theory would have to fit $C(\theta )$ with six free parameters by chance, and we
have shown that up to now [further research is necessary in the fitting of $C(\theta )$] the
standard cosmological model makes it better than would be expected if it were a wrong
theory. That is, the standard model looks correct, otherwise it would not be able to
produce the accurate fit of $C(\theta )$.
If we had found that $f_6$ produced a fit as good as or better than the standard model then it would not
mean that a fit $f_6$ represents a theory which could be an alternative to the standard model, but
we might wonder whether standard cosmology is real or just an instrumental
model to save the data (as in Ptolemaic astronomy), at least with regard to the CMBR anisotropies.
However, this was not the case.

In  cosmologies different from the standard model, there were also attempts to fit the CMB power spectrum with models with  a similar number of free parameters.
Narlikar et al.\cite{Nar03,Nar07} fitted the $C_\ell $ with five--six parameters apart from the amplitude within a model which has nothing to do with the origin of the CMBR in the standard cosmology.
Power spectrum peaks at $\ell \approx 6$ to 10, 180 to 220, and 600 to 900 are shown to be respectively related in their Quasi-Steady State cosmology to curvature effects at the last minimum of the scale factor, clusters, and groups of galaxies. For a  MOND (Modified Newtonian Dynamics) cosmology without cold dark matter, it is also possible with few parameters to fit the power spectrum.\cite{McG04,Ang11}
The standard model still fits the third peak better than our generic polynomial fit or most alternative cosmological scenarios as a function of the same number of free parameters, although there are also alternative scenarios (e.g., Ref. \cite{Ang11}) in which the fit is equally good.

\section{Conclusions}
\label{.concl}

Summing up, any hypothesis which tries to explain the origin of CMBR anisotropies must explain the oscillations of the power spectrum, apart---of 
course---from the black-body emission. 
Here and in Paper I, it has been shown that it is a rather normal characteristic expected from any fluid with  clouds of overdensities that emit/absorb radiation or interact gravitationally with the photons, and with a finite range of sizes and distances for those clouds. Apart from the standard cosmological model, other scenarios may also follow these conditions. The interpretation of ``acoustic'' peaks is just a particular case; peaks in the power spectrum may be generated in scenarios which have nothing to do with oscillations due to gravitational compression.

The angular correlation function is well fitted by a set of polynomial functions with a total of $\approx$6 free parameters,
apart from the amplitude. In the power spectrum, the first two peaks are more or less reproduced with these fits, whereas the third peak and higher-order peaks are not. In this respect, the standard cosmological model, or the Quasi-Steady State Model (both of them using six
free parameters to fit the CMBR power spectrum) have not got too much predictive power to describe the first two peaks properties. Nevertheless, the last WMAP data or ACT data 
have given us the chance to test the standard theory with a good fit of the third peak too, providing greater confidence in the standard model.  

Further research is necessary along these lines in order to see whether fits of the CMBR correlation
function/power spectrum as good as the standard model one can be produced with totally 
different models and a similar number of free parameters.

\

{\bf Acknowledgments:}

Thanks are given to F. Sylos Labini, A. Gabrielli, F. Atrio-Barandela and the anonymous referee for helpful comments. Thanks are given to T. J. Mahoney for proof-reading this paper. WMAP is the result of a partnership
between Princeton University and NASA's Goddard Space Flight Center.
ACT operates in the Chajnantor Science Preserve
in northern Chile under the auspices of the Comisi\'on Nacional de Investigaci\'on 
Cient\'\i fica y Tecnol\'ogica (CONI-CYT).

\

\end{document}